\documentclass[conference]{IEEEtran}
\IEEEoverridecommandlockouts
\usepackage{cite}
\usepackage{amsmath,amssymb,amsfonts}
\usepackage{algorithmic}
\usepackage{graphicx}
\usepackage{textcomp}
\usepackage[bookmarks=false]{hyperref}
\usepackage{xcolor}
\usepackage{authblk}
\def\BibTeX{{\rm B\kern-.05em{\sc i\kern-.025em b}\kern-.08em
    T\kern-.1667em\lower.7ex\hbox{E}\kern-.125emX}}
\begin{document}

\title{A Real-Time Whole Page Personalization Framework for E-Commerce\\}
\author{Aditya Mantha, Anirudha Sundaresan, Shashank Kedia, Yokila Arora, Shubham Gupta, \\Gaoyang Wang, Praveenkumar Kanumala, Stephen Guo, Kannan Achan} \affil{Walmart Labs\\Sunnyvale, CA, USA\\ \{aditya.mantha, anirudha.sundaresan, shashank.kedia, yokila.arora, shubham.gupta0, \\gaoyang.wang, pkanumala, sguo, kachan\}@walmartlabs.com }

\newcommand{\quotes}[1]{``#1''}
\maketitle

\begin{abstract}
E-commerce platforms consistently aim to provide personalized recommendations to drive user engagement, enhance overall user experience, and improve business metrics. Most e-commerce platforms contain multiple carousels on their homepage, each attempting to capture different facets of the shopping experience. Given varied user preferences, optimizing the placement of these carousels is critical for improved user satisfaction. Furthermore, items within a carousel may change dynamically based on sequential user actions, thus necessitating online ranking of carousels. In this work, we present a scalable end-to-end production system to optimally rank item-carousels in real-time on the Walmart online grocery homepage. The proposed system utilizes a novel model that captures the user's affinity for different carousels and their likelihood to interact with previously unseen items. Our system is flexible in design and is easily extendable to settings where page components need to be ranked. We provide the system architecture consisting of a model development phase and an online inference framework. To ensure low-latency, various optimizations across these stages are implemented. We conducted extensive online evaluations to benchmark against the prior experience. In production, our system resulted in an improvement in item discovery, an increase in online engagement, and a significant lift on add-to-carts (ATCs) per visitor on the homepage.
\end{abstract}


\section{Introduction}
Modern e-commerce companies provide personalized user experiences through their homepages. They provide a curated assortment of various heterogeneous components (text, hyperlinks, images, videos, advertisements, etc.), each potentially optimizing for different business metrics. To determine the optimal page layout, numerous eye-tracking studies have been conducted to analyze user behavior as they browse webpages. These studies show a clear positional bias i.e., a user is more likely to browse the top and left parts of a webpage~\cite{granka2004eye,goldberg2002eye}. An e-commerce system that dynamically positions carousels based on user behavior and context is more likely to engage and retain users rather than a static layout~\cite{e-comm-ex}.

Dynamically re-ranking carousels to personalize page layout requires that the system be highly scalable and efficiently incorporate real-time feedback signals, such as user-item interactions and user-carousel interactions. Apart from scalability of the system, layout personalization requires two important considerations. First, there are multiple user-personalized carousels which may individually have their own objectives. These objectives may range from encouraging users to discover previously unseen items to assisting users re-order items easily based on their purchase patterns. This may lead to optimizing the page layout over different, and potentially competing, axes. In our use case, due to business constraints, individual objectives of the carousel need to be honored by preserving the initial ordering of items within the carousel. Second, since a layout optimization system has dependencies upon various upstream item carousel models, some of which may only be inferred at real-time, it is necessary for the layout optimization system to be able to perform carousel ranking inference in an online fashion at low latency. Given these considerations, it makes this layout optimization system uniquely challenging to design. 

In this work, we present a production system that optimizes the ranking of carousels for the Walmart online grocery homepage\footnote{\url{https://www.walmart.com/grocery}}. Our contributions can be summarized as follows:
\begin{itemize}
    \item We introduce an end-to-end scalable production system which provides personalized ranking of carousels for millions of users in real-time. 
    \item We present an offline training and online inference framework that relies on real-time user feedback signals and provides the optimal carousel ranking at low-latency.
    \item We formulate a model to facilitate re-purchase of items and promote discovery of new items based on users' interests, which can be easily extensible to similar settings.
\end{itemize}

\section{Related Work}
Fragmentation of web pages into heterogeneous components provide a platform for rendering different designs based on user preferences~\cite{frag_q1, christos2007web}. There has been significant work in page layout optimization for web search engines~\cite{metrikov2014whole,wang2016beyond,yue2010beyond}. For instance, \cite{wang2018optimizing} proposes a framework for whole page personalization by utilizing content and presentation features to train an ensemble of Quadratic Feature Model and Gradient Boosted Decision Tree Model (GBDT) to predict page-wise user response. The user's interactions with a page can also be modeled as a Markov chain which can be inferred using click logs for optimizing search results presentation~\cite{chierichetti2011optimizing}. 

Component-based ranking systems have also been deployed for optimizing constrained objectives from multiple stakeholders for video recommendations~\cite{ding2019whole} and social networks~\cite{agarwal2015constrained}. Multi-objective optimization problems in personalized click shaping has also been tackled utilizing a Lagrangian duality formulation~\cite{yahoo}. Contextual Bandits provide a natural framework that learns from user behaviour. It learns over time to maintain the required balanced between helping user re-order items (i.e. exploitation) and promoting new items of interest (exploration). It has shown great promise in areas of news article recommendation and ads recommendation\cite{wu2016contextual,tang2015personalized,tang2013automatic}. This framework however, requires a learning phase in which sub-optimal orderings may be shown which may not be appropriate in some business problems. To the best of our knowledge, our work is the first to present an end-to-end system for dynamic ranking of carousels in the online grocery domain.

\section{Problem Statement}
Consider a set of users $\mathcal{U}$ and a set of $N$ personalized carousels for a user $u\in\mathcal{U}$ as $\mathcal{M}_{u}=\{m_{uk} \mid k\in\{1, 2,..., N\}\}$, where $k$ acts as an identifier for a carousel $m_{uk}$. Let $Z ( \le N)$ denote the number of zones where each zone serves as a placement for carousels on the webpage. Each carousel is defined as an immutable ordered list of $M$ items from a global set of items $\mathcal{I}$. Let $\pi_u(i)$ represent the carousel assigned to zone $i$ for user $u$ and $\Pi^{Z}_{u}$ denote the set of all possible assignments of carousels to zones, such that $\pi_u \in \Pi^{Z}_{u}$ is an injective function $\pi_{u}:\{1,2,...,Z\} \to \mathcal{M}_{u}$. Consider a scoring function $\mathcal{F}:\Pi^{Z}_{u} \to \mathcal{R}$ that scores every assignment $\pi_u$, then the objective is to find $\pi_u$ that maximizes $\mathcal{F}( \pi_u )$, without altering the ordering of items within a carousel.

Since $\Pi^{Z}_{u}$ is an exponentially large space, a score-and-sort approach is used. Let $\mathcal{M} = \{ (u, m_{uk}) \mid   u \in \mathcal{U} \wedge m_{uk} \in \mathcal{M}_{u} \}$ denote the set of all possible user-carousel pairs. Given a pointwise scoring function $\phi:\mathcal{M} \to \mathcal{R} $ that maps every carousel for a user to a score, selecting top $Z$ carousels according to $\phi$ produces a valid assignment for the carousels. In real-world scenarios, scoring functions such as $\phi$ are modeled to reflect business requirements.

\section{Model} \label{model}
\subsection{Outline} \label{outline}
In our system, $\phi(u, k)$ (shorthand for $\phi(u, m_{uk})$) is modeled as a linear combination of $\alpha_{uk}$ and $\gamma_{uk}$, which captures the user's affinity to a carousel and the user's propensity to discover new items respectively, as described in~\eqref{AllOver}. Since only the relative ranking of carousels is needed, $\phi(u,k)$ can be reduced to a convex combination with weight $w \in [0,1]$.
\begin{equation}
\phi(u,k) = w\alpha_{uk} + (1-w) \gamma_{uk}
\label{AllOver}
\end{equation}

Further, we have various constraints applied on this optimization problem based on specific business needs. For example, the system might need to consider carousel-specific capacity constraints while assigning carousels to zones on the page. Based on these business requirements, the weight $w$ can be trained.

Since our production system serves millions of users, we use a scalable collaborative filtering based matrix factorization (MF)~\cite{hu2008collaborative} approach to model $\alpha_{uk}$ and $\gamma_{uk}$. Depending on use case, this can be replaced by any generic user-item embedding framework, such as NCF~\cite{he2017neural} and LightGCN~\cite{he2020lightgcn}.

\subsection{User-Carousel Affinity} \label{user_carousel_affinity}
Given a carousel $m_{uk}$ consisting of an immutable ordered list of $M$ items $[i^{uk}_{1}, i^{uk}_{2},...,i^{uk}_{l}, ..., i^{uk}_{M}]$, the probability that a user $u$ will interact with an item $i^{uk}_{l}$ is denoted by $P(i^{uk}_{l}, m_{uk} | u)$. For each user-carousel pair $(u,m_{uk})$, the user-carousel affinity score $\alpha_{uk}$ is assumed to be proportional to the sum of these probabilities weighted by $\omega_l$, as given by:
\begin{equation}
\begin{split}
\alpha_{uk} & \propto \sum^{M}_{l=1} \omega_{l} P(i^{uk}_{l}, m_{uk} | u) \\
 & = P( m_{uk} | u ) \sum^{M}_{l=1} \omega_{l}  P( i^{uk}_{l} | u, m_{uk} )\\
\end{split}
\label{eq:alpha_uk_1}
\end{equation}

where $ P( m_{uk} | u ) $ is the prior probability of the user interacting with a carousel and $P( i^{uk}_{l} | u, m_{uk} ) $ is the probability that the user will purchase an item $i^{uk}_l$ in the carousel.

\subsubsection{User-Carousel Prior}
The prior probability of a user interacting with a given carousel $P( m_{uk} | u )$ is modeled as a beta distribution with parameters $a_{uk}$ and $b_{uk}$, using historical user interaction data. The expected value of this beta distribution is used to estimate the prior probability, denoted by $\lambda_{uk}$, as follows:
\begin{equation}
\lambda_{uk} = \frac{a_{uk}}{a_{uk}+b_{uk}}
\label{beta_prior}
\end{equation}

The user-item interaction events are captured on the e-commerce platform as $(u, m_{uk}, e, t)$, where $t$ is the event timestamp and $e \in \mathcal{E} = \{view, click, ATC\}$, $\mathcal{E}$ being the set of all possible user-carousel interaction events. A event is marked as \quotes{view}, \quotes{click}, or \quotes{ATC} if any of the items in the carousel is viewed, clicked, or added to cart, respectively, by the user in a session. 

Since model inference is performed online, fast and timely updates of user-carousel priors are required to capture the most recent user behaviour. Using a beta distribution to model the prior probability provides the advantage of quick online updates, since the posterior distribution is also a beta distribution with parameters $\hat{a}_{uk}$, $\hat{b}_{uk}$ given by:
\begin{equation}
(\hat{a}_{uk}, \hat{b}_{uk}) \longleftarrow
\begin{cases}
    (a_{uk} + 1, b_{uk}),& \text{if } e =\{click\} \lor \{ATC\} \\
    (a_{uk}, b_{uk} + 1),& \text{if } e = \{view\}  \\
    (a_{uk}, b_{uk}),& \text{otherwise}
\end{cases}
\label{beta_update}
\end{equation}

\subsubsection{User-Item Affinity}
Given a user-item transaction matrix, we train an MF model to learn user and item embeddings and obtain user-item affinity scores $\hat{r}_{ui}: \mathcal{U}\times\mathcal{I} \to \mathcal{R}$ as $\hat{r}_{ui} = x_{u}^Ty_{i}$. Since $\hat{r}_{ui}$ captures a user's affinity towards a particular item and given that item $i^{uk}_{l}$ exists in a carousel $m_{uk}$, we can use $\hat{r}_{ui^{uk}_{l}}$ as a representative score for $P( i^{uk}_{l} | u, m_{uk} )$, thereby $P( i^{uk}_{l} | u, m_{uk} ) \propto \hat{r}_{ui^{uk}_{l}}$. Given $\lambda_{uk} =P( m_{uk} | u )$ and using~\eqref{eq:alpha_uk_1} we get:
\begin{equation}
\alpha_{uk} \propto \lambda_{uk} \sum^{M}_{l=1} \omega_l \hat{r}_{ui^{uk}_{l}}
\label{exploit}
\end{equation}

To capture horizontal position bias observed while browsing webpages~\cite{left_attention}, the weights $\omega_l$ in~\eqref{exploit} are modeled using a logarithmic reduction factor. Without loss of generality, a unit constant of proportionality can be assumed, as we are only concerned with the relative scoring when ranking carousels, thus reducing~\eqref{exploit} to: 
\begin{equation}
\alpha_{uk} = \lambda_{uk}  \sum^{M}_{l=1} \frac{ \hat{r}_{ui^{uk}_{l}} }{log(1+l)} 
\label{exploit_final}
\end{equation}

\subsection{User-Carousel Discovery}\label{discovery_section}
In this section, we develop a model to promote discovery of items by assigning a discovery score $\gamma_{uk}$ to each user-carousel pair. All items in the catalog are divided into a set of categories $\mathcal{C}$, defined by a many-to-one function $\psi:\mathcal{I}\to\mathcal{C}$. The user-category transaction matrix is denser than the user-item transaction matrix, therefore modeling user-category score better captures affinity towards categories the user has not purchased from before. We train an MF model on user-category transaction data to obtain the user-category affinity scores $\hat{s}_{uc}: \mathcal{U}\times\mathcal{C} \to \mathcal{R}$.

For a given user $u$ and category $c \in \mathcal{C}$, the user-category discovery score is defined by a function $g(u,c)$ of $\hat{s}_{uc}$ and $\eta_{uc}$, where $\eta_{uc}$ is the number of times user $u$ has purchased items from category $c$ over a period of time. The more purchases from a category, the higher the chance that a user has explored that category. Hence, to ensure higher scores for unexplored categories, $g(u,c)$ is modeled as $g(u,c) = \hat{s}_{uc}e^{ -\eta_{uc} }$.

For a user $u$ and carousel $m_{uk} = [i^{uk}_{1}, i^{uk}_{2},...,i^{uk}_{l}, ..., i^{uk}_{M}]$,
where item $i^{uk}_l$ is in category $\psi(i^{uk}_{l})$, using a similar logarithmic reduction factor for the ordering as described in Section~\ref{user_carousel_affinity}, the user-carousel discovery score $\gamma_{uk}$ can be computed as:
\begin{equation}
\gamma_{uk} = \sum^{M}_{l=1} \frac{g(u,\psi( i^{uk}_{l} ) )}{log(1+l)} 
\label{explore_final}
\end{equation}

In the next section, we describe the end-to-end system architecture, which combines the user-carousel affinity $\alpha_{uk}$ and discovery scores $\gamma_{uk}$ to generate carousel ranking in real-time using~\eqref{AllOver}.

\section{System Architecture} \label{SystemArchitecture}
\begin{figure*}[tb]
\centering
\includegraphics[width=\linewidth, height=8.5cm]{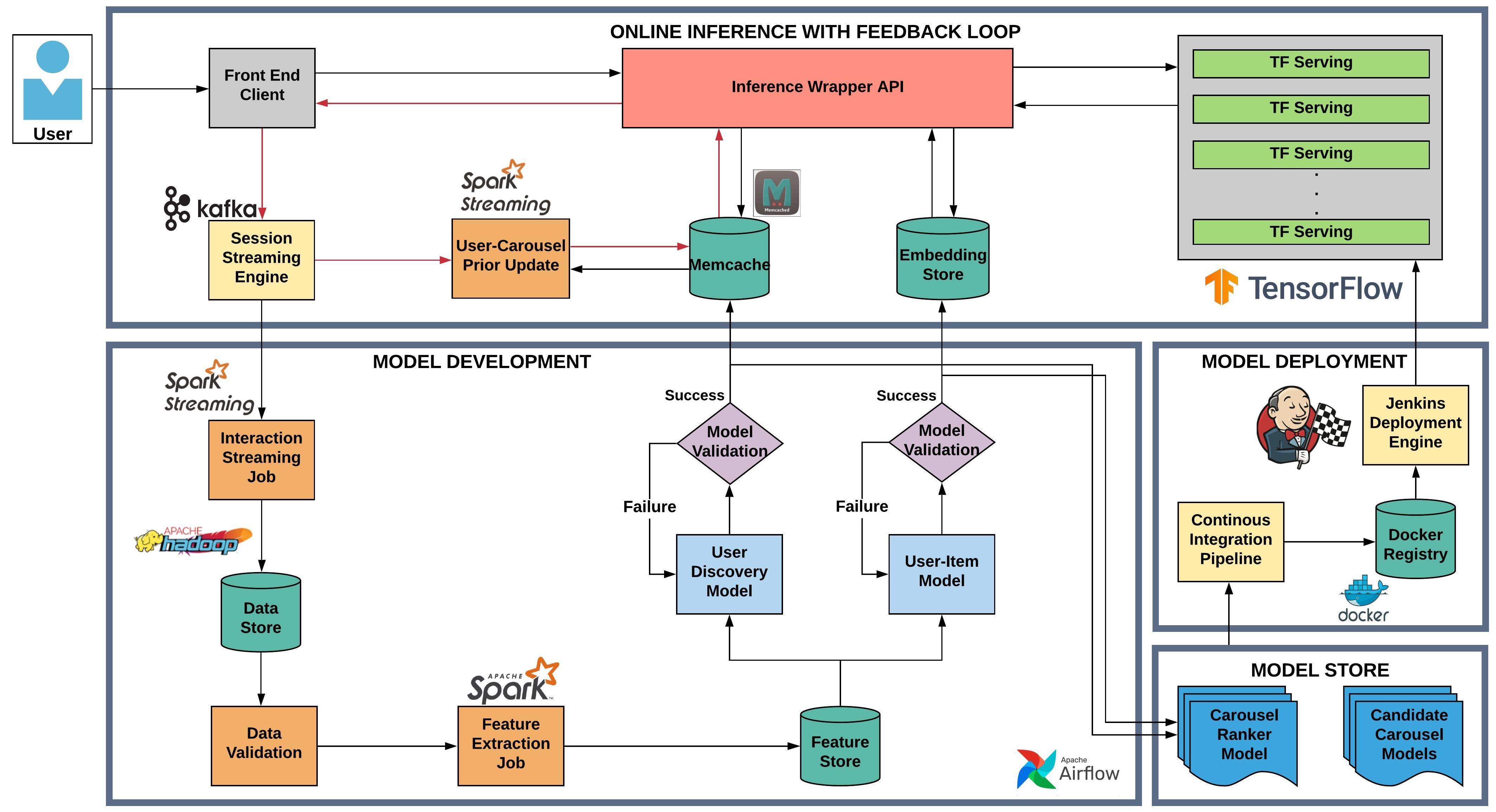}
\caption{System Architecture: Model Development, Model Store and Deployment, Online Inference (with Feedback Loop shown in red)}
\label{fig:SystemArchitecture}
\end{figure*}

We provide an overview of our production system which comprises of three core components: Model Development, Model Store and Deployment, and Online Inference with Feedback Loop, as illustrated in Figure~\ref{fig:SystemArchitecture}.

\subsection{Model Development} \label{offline_arch}
In this section, we explain the components required to develop the user-item and user discovery models. Given the large scale of the Walmart consumer base, we use a high performance Kafka-based distributed event streaming job to store the user-item interaction data in a Hadoop-based distributed data store (HDFS). This HDFS data store serves as the primary central storage and also hosts the item catalog metadata for all items sold on the Walmart online grocery e-commerce platform. After data validation, Beeline and Pyspark workflows are used for feature extraction and subsequent model training on an extensive network of nodes in our distributed computation framework. The extracted user-item transaction features $(u,i,\eta_{ui})$ and user-category transaction features $(u,c,\eta_{uc})$ are stored in the feature store for model ingestion where $\eta_{ui}$ and $\eta_{uc}$ denote the number of transactions for a user-item and user-category pairs respectively.

Using these features, the user-item model (Section \ref{user_carousel_affinity}) and user-discovery model (Section \ref{discovery_section}) are trained. For our use case, we employ transaction data from the past year and obtain 32-dimensional user and item embedding vectors. These learned representations are stored in an embedding store for online inference. The user-category discovery scores $g(u,c)$ are computed and stored in a distributed memory cache. After model training, the model validation phase oversees the outputs of these models and checks for anomalies. Before serving online traffic, using the candidate carousel models with the trained models, the final ranking of carousels of the proposed system is also validated for a subset of users. 

A workflow management platform, Apache Airflow, is used to orchestrate the pipeline stages described above. It enables event-based scheduling and monitoring of complex workflows, along with providing a centralized log monitoring and an alerting system.

\subsection{Model Store and Deployment}
This module consists of a model store, continuous integration (CI) tool, Docker registry, and Jenkins deployment engine. The model store is based on a version control system and serves as a repository of models. When a model is pushed to the model store, a build is triggered by a proprietary CI tool, which then executes automated unit tests, integration tests, and additional checks for model validation. In case of a successful build, a Docker image of the model is generated, pushed to the Docker registry, and packaged as a Docker container. A Jenkins deployment engine then deploys the containerized application to the target production environment, which runs TensorFlow Serving (TF Serving).

\subsection{Online Inference with Feedback Loop}
The online inference system consists of a closed-loop architecture that utilizes real-time user feedback, along with pre-trained model outputs, to dynamically compute the optimal ranking of carousels for each user. The user interacts with the front-end client of the Walmart online grocery homepage through web and mobile interfaces. Two fault-tolerant spark streaming jobs listen to a Kafka stream of user interaction event. One job stores user interactions in the data store as described in Section~\ref{offline_arch}, while the other job periodically updates the user-carousel priors, as per~\eqref{beta_update}. The updated user-carousel priors are stored in a distributed memory cache (Memcached) for high-performance, low-latency, and real-time retrieval. 

When a user visits the homepage, a call is made to the inference-wrapper API to retrieve the optimal ranking of carousels to be displayed. Then, the inference-wrapper API calls TF Serving, where inference is performed for all candidate carousels. TF Serving allows for a unified framework to serve and version these multiple candidate carousel models, while keeping the same server architecture and APIs. Once the candidate carousels are obtained from TF Serving, the inference-wrapper API fetches the required data ($x_u, y_i, \lambda_{uk}, g(u,c)$) for the carousel ranking computation. The embedding store provides the user embedding $x_u$ and the item embeddings $y_i$ for only those items which are part of any of the candidate carousels. In parallel, the updated user-carousel priors $\lambda_{uk}$ for all candidate carousels $m_{uk}$ of a user, and user-category discovery scores $g(u,c)$ for all categories $c$, are obtained from the distributed memory cache. This retrieved data is then passed to TF Serving to compute the user-carousel scores, as described in Section~\ref{model}, according to the following equation: 
\begin{equation}\label{eq:1}
    \phi(u,k) = w\lambda_{uk}\sum^{M}_{l=1} \frac{ x^{T}_{u}y_{i^{uk}_{l}} }{log(1+l)} + (1-w) \sum^{M}_{l=1} \frac{g(u, \psi(i^{uk}_{l}))}{log(1+l)} 
\end{equation}

Using the computed user-carousel scores, the optimal ranking of the candidate carousels is generated by TF Serving and returned to the front-end client to be displayed to the user. Our system combines the benefits of a model training infrastructure, a model deployment framework, and an online inference setup, providing a personalized ranking of carousels to millions of users in real-time at low latency. 

\section{Results}
We conducted a large scale online A/B test on a random sample of users on the Walmart online grocery homepage. In our experimental setup, a pool of 10 carousels were ranked in 10 zones of the homepage. The control group was provided a static ordering of carousels which was based on historical performance of carousels aggregated across all users. On the other hand, the treatment group was presented a dynamic carousel ranking generated by our proposed framework. We observed a lift of 17.79\% ATC per visitor on the homepage, validating the efficacy of our personalized carousel ranking system. We also noted an increase of 0.78\% in item page visits, indicating that our system promotes discovery of previously unseen items. The increased user engagement and item discovery resulted in an overall lift of 0.07\% in all online orders. Our proposed framework thus significantly outperforms the previous system, with an additional latency of only 1.29ms derived from the real-time carousel ranking computations in the TF Serving layer.

\section{Conclusion}
In this work, we present a large-scale production system to personalize the ordering of carousels on the Walmart online grocery homepage. Our system incorporates real-time feedback signals, considers positional biases, and dynamically updates carousel ordering at low latency. We infer user behaviour through the use of an online feedback loop by tracking user interests and updating user-carousel priors in real-time. Our ranking model is modular in design and can be easily extended to other carousel ranking applications. Online experiment results show that our system achieves a significant improvement in user interaction metrics as well as business metrics. The flexible design of our system allows for easy incorporation of additional real-time contextual signals to further improve the system performance. Extending to a contextual multi-armed bandit framework using Thompson sampling on the prior beta distribution is another avenue for exploration.
\bibliography{references}
\bibliographystyle{IEEEtran}

\end{document}